%% file: main.tex
\documentclass{llncs}

\usepackage{natbib}
\usepackage{graphicx}

\usepackage{fancyhdr}
\usepackage{lastpage}

\usepackage{units}

\begin{document}

\title{A short review and primer on cardiovascular signals in human computer interaction applications}
\author{Andreas Henelius}
\institute{Finnish Institute of Occupational Health\\
\email{andreas.henelius@ttl.fi},\\
PO Box 40, Helsinki, 00250, Finland
}

\maketitle

\begin{abstract}
The use of psychophysiologic signals in human-computer interaction is
a growing field with significant potential for future smart
personalised systems. Working in this emerging field requires
comprehension of different physiological signals and analysis
techniques.

Cardiovascular signals such as heart rate variability and blood
pressure variability are commonly used in psychophysiology in order to
investigate phenomena such as mental workload. In this paper we
present a short review of different cardiovascular metrics useful in
the context of human-computer interaction.

This paper aims to serve as a primer for the novice, enabling rapid
familiarisation with the latest core concepts. We emphasise everyday
human-computer interface applications to distinguish from the more
common clinical or sports uses of psychophysiology.

This paper is an extract from a comprehensive review of the entire
field of ambulatory psychophysiology, with 12 similar chapters, plus
application guidelines and systematic review. Any citation to this
paper should be made using the following reference:

{\parshape 1 2cm \dimexpr\linewidth-1cm\relax
B. Cowley, M. Filetti, K. Lukander, J. Torniainen, A. Henelius, L. Ahonen, O. Barral, I. Kosunen, T. Valtonen, M. Huotilainen, N. Ravaja, G. Jacucci. \textit{The Psychophysiology Primer: a guide to methods and a broad review with a focus on human-computer interaction}. Foundations and Trends in Human-Computer Interaction, vol. 9, no. 3-4, pp. 150--307, 2016.\\
\url{http://dx.doi.org/10.1561/1100000065}
\par}


\end{abstract}

\input{ch1_cardiovascular}

\bibliographystyle{abbrvnat_mod}
\bibliography{ch1_cardiovascular_bib}

\end{document}

%% file: ch1_cardiovascular.tex
\section{Introduction}
The heart is innervated by both the sympathetic and the
parasympathetic branch of the ANS. The sympathetic branch, tied to
stress `fight-or-flight' responses, tends to increase heart rate,
whereas parasympathetic activity, representing `rest-and-digest'
behaviour, decreases it. The rate at which the heart beats and
variations thereof hence reflect the activity of the ANS. Accordingly,
various metrics describing the ANS activity can be derived from
cardiovascular signals. For instance, heart rate variability metrics
derived from the electrocardiogram (ECG) are widely used in
psychophysiology \citep{malik:1996:a} -- for example, to investigate
phenomena such as mental workload. The cardiovascular system responds
to sympathetic and parasympathetic activation within a few
seconds \citep{berntson:1997:a}. However, cardiovascular metrics used
in psychophysiology are typically analysed on a time scale of
minutes \citep{malik:1996:a} in the case of short-term heart rate
variability metrics. For instance, the Trier Social Stress
Test \citep{kirschbaum:1993:a} has been used extensively to induce
psychosocial stress, and cardiovascular and endocrine responses are
well documented \citep{kudielka:2007:a}. In this connection, one can
gain insight from a study by \citet{lackner:2010:a} investigating the
time course of cardiovascular responses in relation to mental stress
and orthostatic challenge in the form of passive head tilt-up. There
are also longer rhythms evident in cardiovascular metrics, such as
circadian patterns in heart rate variability \citep{huikuri:1994:a}.

The cardiovascular measurement techniques considered here are (i)
electrocardiography, (ii) blood pressure measurement, and (iii)
photoplethysmography; however, the activity of the heart can also be
measured by means of various other techniques, such as
ballistocardiography \citep{lim:2015:a} or Doppler \citep{lin:1979:a}.

\section{Measurement of cardiovascular signals}
Cardiovascular signals can be measured continuously via non-invasive
techniques and have been widely utilised in the measurement of mental
workload \citep{aasman:1987:a}. The ECG represents the electrical
activity of the heart, and the measurement is carried out with chest
electrodes \citep{malmivuo:1995:a}. From the ECG it is possible to
extract several signals, among them heart rate, or HR (denoting the
absolute pace at which the heart beats) and heart rate variability
(HRV), which is an umbrella concept for all metrics describing how the
rhythm of the heart varies. To record HR and HRV, it is sufficient to
record one lead, as only the R-peaks of the ECG waveform are required;
for details, refer to \citet{berntson:1997:a}. For instance,
affordable sports watches can be used for obtaining a signal suitable
for HRV analysis \citep{gamelin:2006:a}.

Measuring continuous arterial blood pressure (BP) is technically more
demanding than ECG measurement and requires more advanced
equipment. In addition, long-term measurement of BP is not as
unobtrusive as corresponding ECG measurement. Continuous BP can be
measured from the finger, via the method of
Pe\~{n}\'{a}z \citep{penaz:1973:a} as implemented in, for example, the
Finapres device \citep{wesseling:1990:a} and its ambulatory version,
the Portapres.

One can obtain a photoplethysmogram (PPG) either by using methods that
require skin contact or remotely. See, for example,
\citet{allen:2007:a} for a review on the measurement of PPG. In
transmission PPG, the tissue is between the light source and the
receiver, whereas in reflective PPG the light source and the receiver
are next to each other on the surface of the skin, with the light only
bouncing through the tissue from the source to the receiver. The PPG
is typically obtained from the finger and the pinna by transmission
and from the wrist via reflection. The PPG measurement can be
performed remotely without skin contact by using imaging techniques to
consider changes in the pulse \citep{sun:2012:a} (imaging techniques
are discussed in \citet[Section~3.9]{primer2016}). Remote PPG has been
used to study, for example, vasoconstrictive responses during mental
workload by means of an ordinary webcam \citep{bousefsaf:2014:a}. Work
related to this, by \citet{vassend:2005:a}, has used laser Doppler
flowmetry to investigate facial blood flow during cognitive tasks.

The plethysmographic pulse amplitude (PA) depends on the degree of
vasoconstriction, which, in turn, is affected by cognitive
load \citep{iani:2004:a}. The pulse transit time (PTT) in the PPG has
been found to be correlated with BP \citep{ma:2005:a, he:2013:a}.  It
should be noted that vasoconstrictive effects are not visible in the
ear \citep{awad:2001:a}. The use of reflective PPG has become popular
in several consumer sports watches, such as the A360 from Polar
Electro (Kempele, Finland); Forerunner 225 from Garmin Ltd
(Schaffhausen, Switzerland); Charge from Fitbit, Inc. (San Francisco,
CA, USA); Apple Watch from Apple, Inc. (Cupertino, CA, USA); and
Microsoft Band from Microsoft, Inc. (Redmond, WA, USA). Reflective PPG
is used also in research equipment such as the E4 from Empatica
Inc. or various sensors from Shimmer Sensing (Dublin, Ireland).

\section{Methods}
\subsection{Analysis of cardiovascular signals}
In using the cardiovascular signals to investigate the activity of the
autonomic nervous system, it is the variability of the signal that is
of interest. The cardiovascular signals are hence analysed on a
beat-by-beat basis. Raw, continuous cardiovascular signals such as the
ECG, PPG, or continuous beat-by-beat BP signal must therefore be
preprocessed. The goal with the preprocessing is to remove artefacts
from the recorded signals through various methods and reliably convert
the raw signals into event series, where an event corresponds to some
property of one beat of the heart. Accordingly, there are as many
events in the event series as there are heart beats in the raw
signal. Different cardiovascular signals give rise to different event
series. For the ECG, the resulting event series is called an
inter-beat interval (IBI) series or an RR series and each event
corresponds to the duration between consecutive heart beats, typically
measured in milliseconds. The term `RR series' comes from the fact
that the R-peak in the ECG waveform is used as the marker for a heart
beat and each event is the time from one R-peak to the next
R-peak. Similarly, for the PPG the event series is an interpulse
interval time series reflecting changes in blood volume in the tissue,
which varies with the action of the heart. For the continuous BP
signal, it is possible to form, for example, three event series
wherein each event in the respective series corresponds to the
systolic blood pressure (SBP) for each heart beat, the diastolic blood
pressure (DBP), or the mean blood pressure (MBP).

\begin{figure}[ht!]
\centering
\includegraphics[width = \textwidth]{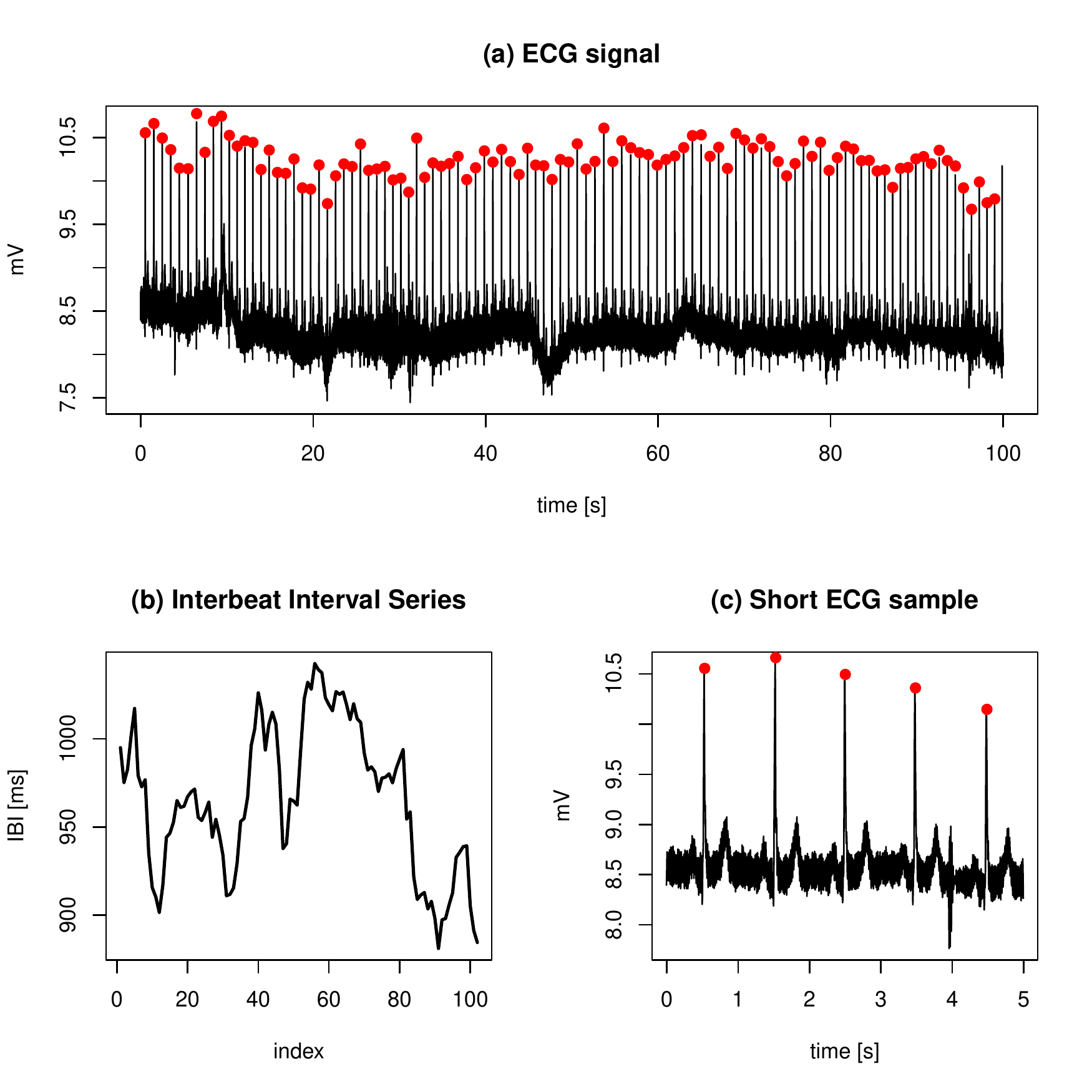}
\caption{Example of the formation of an inter-beat interval (IBI) series from an ECG. The R-peaks in the ECG are shown in red. Panel \textbf{a} shows a 100-second ECG signal, and the corresponding IBI series is shown in panel \textbf{b}. Panel \textbf{c} shows a five-second segment of the ECG signal.}
\label{fig:ecg_hrv}
\end{figure}

Panel \textbf{a} in Figure~\ref{fig:ecg_hrv} shows a 100-second sample
of ECG data from the Physionet \citep{goldberger:2000:a} Fantasia
database \citep{iyengar:1996:a}. R-peaks in the ECG have been
identified and are shown in red. The resulting IBI series is presented
in Panel \textbf{b}, and Panel \textbf{c} shows a shorter, five-second
segment of the ECG signal.

The variability of cardiovascular signals is studied by taking into
consideration several distinct variability metrics calculated from the
event series: HRV from the ECG \citep{malik:1996:a,berntson:1997:a},
blood pressure variability (BPV) from the BP
signal \citep{parati:1995:a}, and pulse rate variability (PRV) from
the PPG \citep{constant:1999:a}. The calculation for these variability
metrics constitutes the main part of the analysis of the
cardiovascular signals. The variability metrics for the individual
signals can be calculated by means of several methods, such as (i)
time-domain, (ii) frequency-domain, and (iii) nonlinear methods. It
should be noted that, although the cardiovascular signals are of a
different nature (e.g., IBIs or SDP values), many of the analysis
techniques developed for HRV analysis are applicable also for BP
analysis \citep{tarvainen:2004:a}. We discuss some HRV metrics
next. Two examples of time-domain measures are the standard deviation
of inter-beat intervals (SDNN), reflecting overall HRV, and the square
root of the mean of the squares of the IBIs (RMSSD), reflecting
short-term HRV \citep{malik:1996:a}. The analysis in the frequency
domain is based on the power spectrum of the IBI signal, derived by
using the Fourier transform, an autoregressive method, or the
Lomb--Scargle method \citep{clifford:2006:a}. In the spectral
analysis, the power of the signal is considered primarily in three
bands: the very low-frequency (VLF) band (\unit[0--0.04]{Hz}), the
low-frequency (LF) band (\unit[0.04--0.15]{Hz}), and the
high-frequency (HF) band (\unit[0.15--0.40]{Hz}). The LF band is
typically linked to sympathetic activation and the HF band to
parasympathetic activation \citep{malik:1996:a}, and the ratio of
power in the LF band to power in the HF band (LF/HF ratio) is used to
describe the degree of sympathovagal balance;
see \citet{billman:2007:a} for a discussion addressing issues related
to the interpretation of the LF/HF ratio.  The nonlinear analysis
methods involve use of metrics such as various entropies.

There are relationships among cardiovascular signals; for instance,
the correlation between HRV metrics derived from fingertip PPG and
from the ECG has, in general, been found to be
high \citep{selvaraj:2008:a,lu:2009:a,lin:2014:a}, although
confounding factors such as respiration should be taken into
account \citep{lee:2010:a}. It ought to be noted that, though HRV and
PRV are related, they are not
identical \citep{constant:1999:a,lu:2009:a,lin:2014:a}. In addition,
factors such as ambient light can affect the PPG signal, and the
latter signal is less stable than the ECG during physical activity.

For a discussion of the analysis of BPV, see \citet{deboer:1985:a}
and \citet{tarvainen:2004:a}. Blood pressure reactivity can be studied
also in terms of baroreflex sensitivity (BRS), which uses a
combination of ECG and BP, as illustrated in other works by,
for example, \citet{mulder:1990:a} and \citet{vanroon:2004:a}. The BRS metric
describes how rapidly the heart rate responds to changes in blood
pressure.

\section{Applications}
The most interesting phenomenon from the perspective of
human--computer interaction is how the various cardiovascular metrics
reflect \emph{mental workload}. It is this aspect of study to which we
direct our focus in the primer.

\subsection{Heart rate variability}
The relationship between mental workload and the ANS response,
reflected in HRV, is complex. However, one can state that HRV
generally is reduced during mental effort, with the degree of
reduction dependent on the level of mental
effort \citep{mulder:1993:a}. It has been shown that average heart
rate is one of the most sensitive metrics for distinguishing between
low and high levels of mental workload in a computerised
task \citep{henelius:2009:a}. In addition, HRV has been used in
occupational settings, a review of which can be found in the work
of \citet{togo:2009:a}.

\citet{garde:2002:a} used HRV to investigate the difference between
two computerised tasks (one using a keyboard, the other using a 
mouse) and concluded that no difference was evident in terms of mental
workload. Differences in HRV metrics were found when the setting featured a
physically demanding computer task. In a study 
\citep{hjortskov:2004:a} investigating differences in a computerised
task with different difficulty levels, the researchers found
differences in the HRV LF/HF ratio between the tasks. Hjortskov and
colleagues also concluded that HRV is more sensitive than BPV is to
mental stress levels. In a recent study, \citet{taelman:2011:a} found
several HRV metrics (average normal-to-normal interval length, SDNN,
RMSSD, pNN50, LF, and HF) to be affected by the extent of
mental load.

A study by \citet{cinaz:2013:a} investigated the use of HRV for
classifying levels of workload during office work. They found that
RMSSD and pNN50 decreased with the degree of workload experienced,
while the LF/HF ratio increased

In an example of particular frequency bands within the HRV
spectrum, \citet{nickel:2003:a} investigated the 0.1 Hz component of
HRV during the performance of a battery of computerised stress
tests. They concluded that this particular frequency component was not
sensitive to workload level.

The heart pre-ejection period (PEP)
\citep{backs:2000:a,miyake:2009:a} and the T-wave amplitude
(TWA) \citep{myrtek:1994:a,vincent:1996:a} too have been linked to
mental stress. See the work of Lien and colleagues \citet{lien:2015:a}
for a study comparing these indices as metrics of sympathetic nervous
system activity in ambulatory settings.

Heart rate variability has been studied extensively in connection with
measuring the task difficulty or mental workload experienced by
pilots \citep{jorna:1993:a,roscoe:1992:a,roscoe:1993:a}. In addition,
research by \citet{wilson:2002:a} investigated various
psychophysiological measurements during flight and found that HR was
more sensitive than HRV to task difficulty.

\subsection{Blood pressure}
Measurements of BP, BPV, and BRS have been applied in multiple studies
related to mental workload.

For instance, in a study \citep{stuiver:2014:a} that considered two
simulated tasks (ambulance dispatching and a driving simulator), the
researchers found that HR increased during the dispatch task; BRS
decreased; and BP showed an initial increase, after which it continued
to rise, albeit more slowly. For the driving task, BP initially
increased but then fell to near baseline levels. Both BRS and HR
decreased during the task. The researchers concluded that there are
task-specific differences that lead to different types of autonomic
control.

A study carried out by \citet{madden:1995:a} utilised a computerised quiz to
induce mental stress. It was found that BPV decreased and systolic BP
rose as the degree of mental stress increased. Similar results
were obtained in another study \citep{ring:2002:a}, in which
mental stress was induced by mental arithmetic, leading to an increase
in mean BP.

Finally, a study by \citet{robbe:1987:a} investigated BRS during
mental arithmetic and memory tasks. Its conclusion was that the
modulus (the gain between BPV and
HRV \citep{mulder:1988:a,robbe:1987:a}) decreased during task
performance. Additionally, blood pressure was recorded during flight
in research conducted by \citet{veltman:1998:a}, and the modulus was
found to be a good index for mental effort.

\subsection{Photoplethysmography}
In addition to the measures discussed above that reflect various
aspects of cardiac variability, the autonomic activity can be studied
by observing vasoconstriction. As noted above, mental stress is
reflected in peripheral vasoconstriction, which is visible in the PPG
signals as a decreased pulse amplitude (PA). For
instance, \citet{iani:2004:a} investigated the peripheral arterial
tone (measured by means of a pneumatic plethysmograph) during the
performance of a computerised memory task. They found that subjects
exhibited vasoconstriction during the more demanding memory tasks in
their experiment. Similar results were found in an experiment
involving simulated flight, wherein vasoconstriction was observed
during difficult phases of the simulation \citep{iani:2007:a}.

Pulse rate variability can also be analysed in a  fashion similar to HRV.
\citet{yoo:2011:a} examined detection of mental stress by using the
PPG signal and PRV, while other researchers \citep{yashima:2006:a,
kageyama:2007:a} have explored the use of wavelet analysis for mental
stress detection via the PPG signal. \citet{arai:2012:a} estimated
mental stress by considering the LF/HF ratio calculated from the PPG
signal, and the resulting extracted mental stress metric was used to
control the functioning of a mail program on a smartphone.

\section{Conclusions}
Cardiovascular signals and the variability metrics extracted from them
are used extensively for the determination of mental workload during
various types of task performance. Of the signals considered in this
section of the paper, the ECG and PPG signals are easy to measure, and
good-quality recordings can be obtained with affordable devices such
as sports watches. The recent incorporation of PPG sensors into
several wrist-worn devices means that such devices could well be
usable for long-term measurements. The use of remote PPG techniques,
of various types, that employ ordinary webcams might be highly
suitable for computer work. Blood pressure, however, is not well
suited to the prolonged measurement of cardiovascular activity at
present, on account of the technical requirements. In terms of
usability, ECG and PPG measurements are hence currently more suitable
for human--computer interaction applications than is BP.

The metrics used for analysing cardiovascular signals are well
established, as illustrated, for example,
by \citet{malik:1996:a}. Furthermore, other metrics describing
variations in cardiovascular signals are being studied, such as the
fractal dimension of the HRV signal \citep{nakamura:1993:a,
caceres:2004:a}.

Cardiovascular metrics have been applied extensively for determining
and monitoring levels of mental workload. However, it appears that
these metrics have not been used thus far for the purpose of adapting
user interfaces to the degree of mental workload. This is a new,
relatively unexplored area.